\documentclass[11pt,twoside]{article}
\usepackage{./asp2014}
\usepackage{natbib}

\newcommand{\pasa}{PASA}

\aspSuppressVolSlug
\resetcounters

\begin{document}

\title{Science with an ngVLA: Local Constraints on Supermassive Black Hole Seeds}
\author{Richard~M.~Plotkin$^1$ and Amy~E.~Reines$^2$
\affil{$^1$International Centre for Radio Astronomy Research - Curtin University, GPO Box U1987, Perth, WA 6845, Australia; \email{richard.plotkin@curtin.edu.au}}
\affil{$^2$Department of Physics, Montana State University, Bozeman, MT 59717, USA; \email{amy.reines@montana.edu}}
}

\paperauthor{Richard~M.~Plotkin}{richard.plotkin@curtin.edu.au}{0000-0002-7092-0326}{International Centre for Radio Astronomy Research }{Curtin University}{Perth}{WA}{6845}{Australia}
\paperauthor{Amy~E.~Reines}{amy.reines@montana.edu}{0000-0001-7158-614X}{Montana State University}{Department of Physics}{Bozeman}{MT}{59717}{USA}

\begin{abstract}
Determining the mechanisms that formed and grew the first supermassive black holes is one of top priorities in extragalactic astrophysics.  Observational clues can be inferred from the demographics of massive black holes (in the ten thousand through million Solar mass range) in nearby low-mass galaxies.  This chapter of the next generation Very Large Array (ngVLA) Science Book describes how an ngVLA  can play a prominent role in developing large samples of weakly accreting active galactic nuclei in low-mass galaxies (out to nearly 1 Gpc), which will help constrain the types of objects that originally seeded the growth of supermassive black holes.
\end{abstract}

\section{Introduction}
Determining the formation mechanism(s) of the first black holes is one of the highest priority research areas in extragalactic astrophysics today.  In this chapter we describe how  a next generation Very Large Array (ngVLA) will provide key new constraints on the types of objects that originally ``seeded'' the growth of supermassive black holes.  In particular, when combined with other multiwavelength probes, an ngVLA will be a crucial player in  performing population studies of black holes in the $10^4 \lesssim M_{\rm BH} \lesssim 10^6\,M_\odot$ range, objects that we will  refer to  as massive black holes (mBHs).

Radio observations provide a window into one of the most energetically important components of accreting black holes, their relativistic jets.  Jets can carry  large amounts of mechanical power away from a black hole accretion disk, making them one of the primary channels through which black holes can influence their large-scale (kpc) environments \citep[e.g.,][]{fabian12}.  As strong synchrotron emitters, jets release large amounts of radiative power in the radio waveband.  Thus,   jetted radio emission  can serve as a signpost that signals the presence of  an accreting black hole.   With its order of magnitude improvement in sensitivity and exquisite spatial resolution, an ngVLA will be a powerful machine for identifying and studying accreting mBHs, including studies on how mBHs influence their environments (for details on the latter, see the chapter by Nyland \& Alatalo on \textit{SMBH Feedback and in Nearby Low-mass Galaxies} in Part VI of this book).  Combining ngVLA radio signatures  with radiation at other wavebands (e.g., X-ray spectral and timing studies, high spatial-resolution infrared/optical spectroscopy and imaging, etc.) will provide a significantly improved  understanding on the local mBH population,  on mBH accretion and feedback, and on how black hole growth was seeded in the early Universe.

\section{The First Black Holes}

 The existence of high-redshift quasars powered by  $\sim$$10^9\,M_\odot$  black holes at $z \gtrsim 7$ suggests that the first black hole seeds were `heavy', with masses $M_{\rm BH} \sim 10^5\,M_\odot$ (which could have formed  from the direct collapse of large clouds of gas, e.g., \citealt{loeb94, begelman06}).  Otherwise, $10^9\,M_\odot$  black holes could not have grown so massive  only 700 Myr after the Big Bang (assuming growth through Eddington limited accretion, e.g., \citealt{mortlock11, banados18}).    However, whether or not $10^5\,M_\odot$ seeds are the exception or the rule remains uncertain, and it is possible that some fraction of  black holes could have been seeded with lighter objects in the $\sim10^2\,M_\odot$ range (i.e., remnants from Population III stars, e.g,. \citealt{haiman01, madau01, madau14}).   
 
 While observations of high-redshift quasars  provide important constraints and boundary conditions, it is  not  feasible with current facilities to directly observe black holes as small as $10^5\,M_\odot$ in the very high-redshift Universe \citep{volonteri16}.    To study  low-mass black holes,  we instead rely on clues that are embedded within  black hole populations found in nearby galaxies.  In particular, the fraction of  galaxies hosting a nuclear black hole  (as a function of galaxy stellar mass) and  black hole/host galaxy scaling relations at low masses  provide local diagnostics on the seed black hole population (see, e.g., \citealt{volonteri10, greene12, reines16a, mezcua17} for  reviews).  

\subsection{Low-mass Black Holes  in Nearby Dwarf Galaxies}
Because of their relatively quiet evolutionary histories, and the ability for supernova feedback to stunt black hole growth, we do not expect black holes in nearby dwarf galaxies\footnote{Following \citet{reines13}, we define dwarf galaxies as having stellar masses $M_\star < 3 \times 10^9\,M_\odot$, which is comparable to the Large Magellanic Cloud.} 
 to have grown much within a Hubble time \citep[e.g.,][]{dubois15, habouzit17}.  In turn, characterizing mBHs in dwarf galaxies  provides a powerful lever arm for constraining the seed black hole population \citep[e.g.,][]{volonteri09, van-wassenhove10, bellovary11}.   In the past decade we have witnessed an explosion in sample sizes of mBHs in dwarf galaxies, growing from a handful of isolated cases 10-15 years ago \citep[e.g.,][]{filippenko03, barth04}, to now homogeneously selected samples reaching $\sim$$10^2$ objects in dwarfs \citep[e.g.,][]{reines13, mezcua18} and other low-mass galaxies \citep[e.g.,][]{greene04, greene07, dong12}.    mBHs in dwarf galaxies typically have masses in the $10^4-10^6\,M_\odot$ range, with the smallest mBH discovered so far residing in the dwarf galaxy RGG 118, weighing in at only $5\times10^4\,M_\odot$ \citep{baldassare15}.  

Discovering mBHs in dwarf galaxies relies on exploiting radiative signatures from accreting mBHs (see \citealt{reines16a} for a recent review).  That is, we are sensitive only to mBHs that happen to be  shining as active galactic nuclei (AGN).    Arguably the most efficient discovery method so far has been based on optical signatures using emission line diagnostics (utlizing data from  large-scale spectroscopic surveys; \citealt{reines13}).\footnote{Infrared searches have also yielded large sample sizes, although see \citet{hainline16} for a discussion on the purity of infrared selected samples.}  
Optical selection, however, is biased toward AGN with  low extinction that are accreting relatively rapidly ($L_{\rm bol}/L_{\rm Edd} > 0.1$).  

Low-mass AGN can also be found through X-ray surveys \citep[e.g.,][]{kamizasa12, schramm13, lemons15, mezcua16, mezcua18}, although X-ray selection is  biased against highly obscured systems. Also, a practical limitation with (current) X-ray surveys  is that economically surveying large areas of sky generally requires  single-epoch X-ray `snapshots' with our most sensitive facilities.  These types of X-ray observations rarely provide enough photons to pursue  high signal-to-noise spectroscopic or timing studies.  Without  spectral and timing information, current X-ray facilities are not usually capable (on their own) of differentiating between accreting mBHs and luminous X-ray binary systems in cases with X-ray luminosities $L_X \lesssim 10^{40} {\rm erg\,s}^{-1}$, except for in a handful of bright systems with long X-ray exposure times, or in some cases with a sufficient number of epochs  to monitor  variability.

\section{Revealing Low-mass AGN with an ngVLA}
\label{sec:fp}
Some of the current practical limitations to pursuing large population studies can  be overcome  by exploiting jetted radio emission from accreting mBHs.  Radio observations have already discovered  a few mBHs in dwarfs \citep{reines11, reines14}, and a new  survey of dwarf galaxies with the Jansky Very Large Array (VLA) is revealing  many new candidates (Reines et al.\ in prep).  As described below, once radio facilities start reaching ngVLA sensitivities, mBH AGN selection can become  efficient through combined multiwavelength efforts. 

At low Eddington ratios ($<$$10^{-2}\,L_{\rm Edd}$), all AGN are believed to launch compact, partially self-absorbed synchrotron jets that will appear point-like even at ngVLA resolutions \citep{blandford79, ho08}.  In this weak accretion regime, the ratio of radio:X-ray luminosity scales in a predictable way with the mass of the accreting black hole according to the fundamental plane of black hole activity,  $\log L_{\rm R} = 0.6 \log L_{\rm X} + 0.78 \log M_{\rm BH} + 7.33$  \citep[e.g.,][]{merloni03, falcke04, plotkin12}, where $L_{\rm R}$ is at 5 GHz, $L_{\rm X}$ is from 2-10 keV, and $M_{\rm BH}$ is in Solar units. 

 Since radio jets are expected to be ubiquitous at low accretion rates, a radio survey can be effective at recovering populations of mBH candidates if undertaken with a sensitive enough telescope.  Advantages of a radio survey include: 
 \begin{itemize}
 \item targeting mBHs in a weak accretion regime will recover objects missed by optical surveys (where the latter rely on diagnostics that are expected to  be  present mostly at higher accretion rates).  Note that for any sensible luminosity function, the majority of AGN accrete at low Eddington ratios;
 \item jetted radio emission will be detectable  from highly absorbed AGN, thereby allowing radio selection to uncover some objects that would be missed by complementary X-ray and/or optical approaches;
 \item at Mpc distances, contamination in the radio waveband from X-ray binary systems should be negligible.  Except,  at ngVLA sensitivities ($5 \sigma_{\rm rms} \approx 1\, \mu$Jy) it is plausible to (rarely) detect emission from  transient radio flares produced by accreting stellar mass black holes out to $\approx$30 Mpc.\footnote{We estimate this limit by  scaling  an extreme 20 Jy radio flare from the Galactic black hole X-ray binary Cyg X-3 \citep{corbel12}.}  

 \end{itemize}
 We note that, in this low-luminosity regime, radio surveys will have to contend with radio emission from  star formation (i.e., free-free emission in \ion{H}{ii} regions) and from supernova remnants.  However, jetted radio emission from weakly accreting mBHs will be compact and unresolved at Mpc distances.  Thus, with long ngVLA baselines ($\sim$$10^3$ km), point-like radio emission would be a strong indicator of a viable AGN candidate \citep[e.g.,][]{reines12}.

Figure \ref{fig:sens} displays expected radio flux densities  at  8 GHz for $10^{-5} < L_{\rm X}/L_{\rm Edd} <  10^{-3}$ AGNs powered by $10^4$ (red swath), $10^5$ (blue swath) and $10^6\,M_\odot$ (grey swath) mBHs as a function of distance.\footnote{We note that we expect all AGN accreting at $\lesssim10^{-2}\,L_{\rm Edd}$ to emit compact radio emission.  We adopt the range $10^{-5}-10^{-3}\,L_{\rm Edd}$ in Figure~\ref{fig:sens} simply for illustrative purposes, to keep the figure from appearing too cluttered.}  
  We also show the current VLA 5$\sigma_{\rm rms}$ detection limit (dotted horizontal line) and an ngVLA 5$\sigma_{\rm rms}$ detection limit (dashed horizontal line) assuming 1 hour integrations. The shaded regions require 2-10 keV X-ray fluxes $F_{\rm X} >10^{-15}\,{\rm erg\,s^{-1}\,cm^{-2}}$ to illustrate the types of AGN that can be accessed with an ngVLA combined with an X-ray facility like \textit{Chandra}.  An ngVLA would be able to detect relatively massive mBHs (10$^6\,M_\odot$) out to nearly 1 Gpc (almost a factor of four farther than accessible by the current VLA within reasonable exposure times).   Excitingly, an ngVLA could detect jetted radio emission from  $10^4\,M_\odot$ mBHs accreting at $L_{\rm X} \gtrsim10^{-4}\,L_{\rm Edd}$, or $10^5\,M_\odot$ mBHs at $\gtrsim 10^{-6}\,L_{\rm Edd}$, at the distance of the Virgo cluster (16.4 Mpc; corresponding to 2-10 keV X-ray fluxes $F_{\rm X} \gtrsim 4 \times 10^{-15}$  and $\gtrsim 4 \times 10^{-16}$ erg s$^{-1}$ cm$^{-2}$, respectively), which we expand on in Section \ref{sec:mw}.

\articlefigure{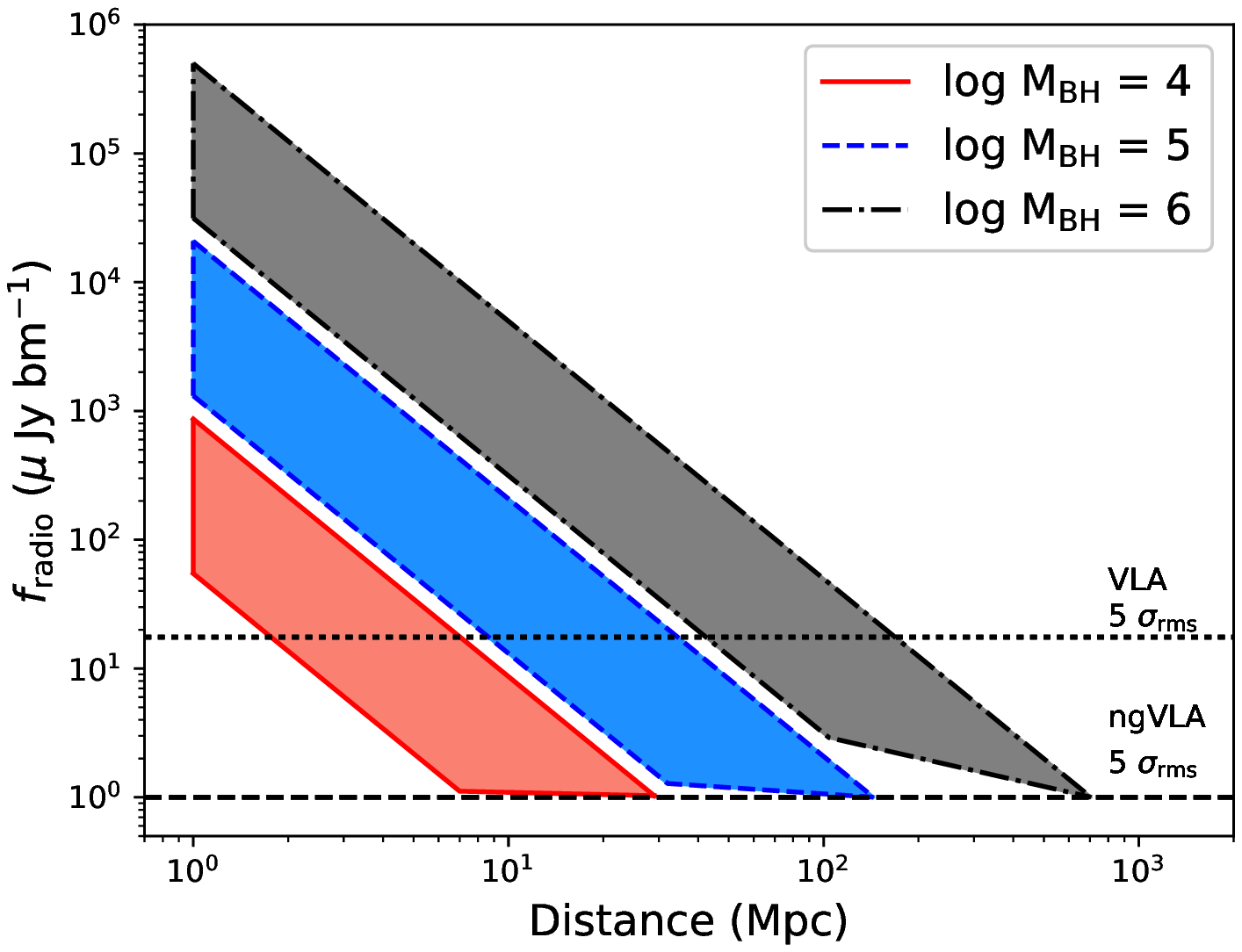}{}{Distances to which an ngVLA could detect compact radio emission from an accreting mBH with 1 hour integrations, if the mBH falls on the fundamental plane of black hole activity \citep{merloni03}, and if the mBH has a 2-10 keV X-ray flux $F_{\rm X} > 10^{-15}$ erg s$^{-1}$ cm$^{-2}$.  The red shaded region bounded by red solid lines illustrates the expected radio flux density (at 8 GHz) for a $10^4\,M_\odot$ mBH accreting between $10^{-5} < L_X < 10^{-3}\,L_{\rm Edd}$; the blue shaded region (bounded by dashed lines) and the grey shaded region (bounded by dashed-dotted lines) illustrate the same for a 10$^5$ and 10$^6\,M_\odot$ mBH, respectively.   5$\sigma_{\rm rms}$ radio detection limits for an ngVLA and for the VLA are shown as dashed and dotted lines, respectively (both assuming 1 hour on source).  An ngVLA could detect relatively massive mBHs out to nearly 1 Gpc, and it could detect mBHs as small as 10$^4\,M_\odot$  at the distance of the Virgo cluster of galaxies ($\sim$$16.4$ Mpc).  Note that the \citet{merloni03} fundamental plane has an intrinsic scatter of 0.88 dex in radio luminosity, which is not included in this figure.
 \label{fig:sens}}

 \subsection{Very Long Baseline Capabilities}
 
The inclusion of very long baselines ($10^4$ km) to an ngVLA would provide an exciting opportunity to measure proper motions from nearby AGN.   We generally expect compact radio emission in the weak accretion regime to remain unresolved at GHz frequencies, even with very long baselines, such that we can expect $\approx$10 micro-arcsec positional accuracy for bright  sources.  With such accuracy, we could measure a radio-emitting mBH in Virgo move by 5$\sigma$ (i.e., $\approx$50 micro-arcsec) in a time baseline of only four years (if we assume a  projected velocity of $\approx$1000 km s$^{-1}$ at 16.4 Mpc, which is a typical peculiar velocity in Virgo).  Over 10-25 years it would be feasible to measure proper motions out to 50 Mpc, for velocities from 500-1000 km s$^{-1}$.  Such precision astrometry would allow us to employ AGN proper motion measurements (in conjunction with multiwavelength radiative signatures) for distinguishing nearby mBHs from more distant AGN.  In some lucky and nearby cases, there is even the potential to measure mBH motion from recoil kicks imparted by gravitational wave emission after a merger.

\section{Jet Physics and Mass Scaling}
We describe in the next section one way in which  revealing new samples of mBHs through an ngVLA would observationally constrain the masses of black hole seeds.   In this section, we  stress that more sensitive radio observations would also allow studies on the physics of jets in the low-accretion rate and low-mass regime.  For example, the sensitivity of an ngVLA would open up economic time-domain studies for bright sources, and its wideband frequency capabilities  would allow radio spectral studies (up to $10^2$ GHz) for an unprecedented large sample of mBHs (which might even allow detailed decomposition of mBH radio jets from radio emission produced by star formation for the brightest objects).  An ngVLA would help constrain the role of AGN feedback in dwarf galaxies and at low accretion rates in the so-called `kinetic' feedback mode  \citep[e.g.,][]{xie17, dashyan18}.  By populating observations of extragalactic jets launched by black holes in the $10^4-10^6 M_{\rm BH}$ range, an ngVLA would also provide empirical constraints to drive studies on the scalability of black hole accretion and jet launching, through comparisons of mBHs to more massive black holes ($\gtrsim$$10^6\,M_\odot$) at the centers of large galaxies,  and to lower-mass stellar black holes  ($\sim$$10\,M_\odot$) in X-ray binary systems.

\section{Multiwavelength Studies of  Black Hole Seeds}
\label{sec:mw}

\subsection{Multiwavelength Coordination}
Constraining  black hole seeds through  AGN  in dwarf galaxies is a multiwavelength endeavor, and it is not a problem that can be tackled looking within only a single band of the electromagnetic spectrum.  However, in a multiwavelength context,  we expect a radio facility like the ngVLA  to play a prominent role.   For example, as discussed earlier, optical selection so far has proved to be effective at higher Eddington ratios ($\gtrsim 0.1 L_{\rm Edd}$), but optical surveys  largely miss  the types of weakly accreting objects for which combined radio/X-ray searches are optimal.  It is not obvious whether (in the near future) large high-spatial resolution optical telescopes will outperform radio surveys in the weak accretion regime:   weakly accreting AGN do not always have obvious optical counterparts in dwarf galaxies \citep[e.g.,][]{reines11, reines14}, and optical AGN diagnostics, like broad emission lines and narrow high-excitation  lines, are expected to become substantially weaker at lower Eddington ratios  \citep[e.g.,][and references therein]{padovani17}.  Thus, efficient optical selection of weakly accreting mBHs would require routine detections of low-excitation narrow AGN emission lines, a method that is possible in principle  but has not yet revealed large numbers of AGN candidates in dwarf galaxies.   Combining radio and optical samples will therefore reveal AGN over a wider range of Eddington ratios than can be  recovered  individually.

It is challenging for current (and near-future) X-ray facilities to economically survey large numbers of dwarf galaxies, especially with sufficient sensitivity to unambiguously control for contamination from X-ray binaries.    Advantages of an ngVLA are that it can \textit{efficiently} reach useful sensitivities in relatively short ($<$1 hour) exposures, and contamination from X-ray binaries will be rare for nearby galaxies, and negligible at $\gtrsim$30 Mpc.  Furthermore,  with its long baselines,  contamination from star formation can be minimised by selecting only point-like radio  emission.

Radio-selected AGN candidates would then await confirmation by other telescopes, including high spatial-resolution X-ray facilities like the \textit{Chandra X-ray Observatory} or the \textit{Athena X-ray Observatory} (Figure~\ref{fig:sens} demonstrates the discovery space that is currently feasible via  \textit{Chandra} followup to ngVLA AGN candidates).  Some of the X-ray brighter systems (e.g., $F_{\rm X} \gtrsim 10^{-13}$ erg s$^{-1}$ cm$^{-2}$) will also appear in future all sky surveys (e.g.,  eROSITA on the \textit{Spectrum-Roentgen-Gamma} satellite).  High spatial-resolution infrared spectroscopy, e.g., with the \textit{James Webb Space Telescope (JWST)}, may also provide  line diagnostics to confirm weakly accreting and/or optically obscured AGN  \cite[e.g.,][]{satyapal09}.

\subsection{Local Constraints on Black Hole Growth}
Identifying  AGN in dwarf galaxies is still a young field observationally.  The progression of connecting observations of mBHs to theoretical models of seed black hole formation will require continuing to systematically assemble even larger samples  over the next decades.  As an example, consider the AGN Multiwavelength Survey of Early-Type Galaxies \citep[AMUSE;][]{gallo08, miller12}, which  performed a \textit{Chandra} X-ray survey of 100 early-type galaxies in the Virgo cluster and 100 non-cluster galaxies at comparable distances to Virgo.  By considering the fraction of nuclear X-ray detections within the 200 object AMUSE sample, \citet{miller15} provide a framework demonstrating how black hole occupation (and seed formation) can be constrained through local observations of weakly accreting AGN.   They place a limit that $>$20\% of local low-mass galaxies ($M_\star < 10^{10} M_\odot$) are occupied by an mBH, and they  cannot rule out full occupation.  Amongst the results from \citet{miller15} we note:
\begin{itemize}
\item their 200 object sample is not large enough  to place stringent constraints on occupation (although see their Figure 4 for how  uncertainties  improve with  sample size); 
\item  with the types of X-ray sensitivities achievable if aiming to pursue $N\sim10^2$ sample sizes,  contamination from X-ray binaries is an important systematic to be controlled for. 
\end{itemize} 
Despite the above, we stress that quantitative methods are already fully developed by \citet{miller15}, who demonstrate that local calculations of the occupation fraction are in reach.  An ngVLA would help improve  sample sizes, and it would help provide confidence that contamination from X-ray binaries is negligible.

Since one  expects full occupation at the high galaxy (stellar) mass end regardless of the seed black hole mass distribution, diagnosing the seed formation channels will require focusing on lower-mass galaxies \citep{miller15}.    A typical \textit{Chandra} X-ray exposure time utilized by the AMUSE survey at the distance of Virgo was $\sim$1.4 h, which allowed identification of a $10^5\,M_\odot$ mBH accreting at nearly $10^{-5}\,L_{\rm Edd}$ (a regime that is just barely accessible with the current VLA; see Figure~\ref{fig:sens}).   With a comparable or shorter time investment on an ngVLA (depending on  overheads), radio observations could identify a $10^5\,M_\odot$ mBH accreting an order of magnitude more weakly ($10^{-6}\,L_{\rm Edd}$), which would clearly improve the sample size (the corresponding X-ray flux would be $\approx 4 \times 10^{-16}$ erg s$^{-1}$ cm$^{-2}$, which is feasible given that only a subset of objects would be so faint).    Thus, identification of  AGN in dwarf galaxies (at ngVLA sensitivites) can be expected to place significantly improved constraints on the local black hole occupation fraction and mBH mass distribution, thereby helping to place observational constraints on the population of objects that seeded the growth of black holes in the early Universe.

\end{document}